\documentstyle[preprint,aps]{revtex}
\tighten
\draft
\widetext
\input{epsf.sty}
\preprint{CLNS 01/1770}
\font\blackboard=msbm10 
\font\blackboards=msbm7 \font\blackboardss=msbm5
\newfam\black \textfont\black=\blackboard
\scriptfont\black=\blackboards \scriptscriptfont\black=\blackboardss

\def\be{\begin{equation}}
\def\ee{\end{equation}}
\def\baray{\begin{eqnarray}}
\def\earay{\end{eqnarray}}

\begin{document}
\title{Comment on 4D Lorentz invariance violations in the brane-world}

\medskip

\author{Horace Stoica}

\medskip

\address{Laboratory for Nuclear Studies \\
Cornell University, Ithaca, NY 14853\\
e-mail:\hspace{6pt} fhs3@mail.lns.cornell.edu}
\medskip
\date{\today}
\maketitle

\begin{abstract}
The brane-world scenario offers the possibility for signals to travel outside our visible 
universe and reenter it. We find the condition for a signal emitted from the brane to return to the 
brane. We study the propagation of such signals and show that, as seen by a 4D observer,
these signals arrive earlier than light traveling along the brane. We also study the horizon problem
and find that, while the bulk signals can travel far enough to homogenize the visible universe,
it is unlikely that they have a significant effect since they are redshifted in the gravitational field of 
the bulk black hole.  
\end{abstract}

\section{Introduction}
The brane-world scenario offers possible solutions to the hierarchy problem
\cite{RS1} and it predicts new effects, like modifications of Newton's force law at short distances
\cite{RS2}, modifications of the cosmological evolution at early times 
\cite{Csaki_0,Binetruy,Eanna,Ellwanger,io,Gregory} and the existence of ``faster-than-light'' gravitational signals
\cite{Kraus,Grojean,Csaki_1,Csaki_2,Caldwell} . The last effect is due to the 
fact that, while the Standard Model fields (namely the photons) are confined to live 
only inside the brane world-volume, gravity can see the whole higher-dimensional space-time. 
Consequently, a ``shorter'' path  that leaves the brane, propagates through the bulk, and re-enters the 
brane later may exist for gravitons. 

This effect was pointed out in Ref.\cite{Halevi}, where the authors argued that all points in the visible
universe can be connected in an arbitrarily short amount of time. The effect was also used in Ref.\cite{Freese} 
to give a solution to the horizon problem without using inflation. In Ref.\cite{Kalbermann} the author argues
that however, closed time-like curves do not form, and therefore causality is maintained.  
The causal structure and the horizon problem in the absence of inflation are also studied in 
Ref.\cite{Ishihara},
and the holographic interpretation of the Lorentz invariance violations is given in Ref.\cite{Creminelli}.

Since the 5D space-time is usually chosen to be AdS-Schwarzschild (AdSS) or AdS-Reissner-N\"ordstrom 
(AdS-RN) in order to localize gravity and reproduce Newton's law in the brane, the 5D metric has the general 
form of Ref.\cite{Birmingham} :
\be
\label{AdDSS_metric}
ds^2=-f\left(r\right)dt^2+r^2d\Sigma_k^2+\frac{dr^2}{f\left(r\right)}
\ee
The 4D universe is seen as a 3-brane moving through this background. Since there is no preferred orientation 
of the brane, there is no reason to expect Lorentz invariance on the brane. The problem is addressed in Ref.
\cite{Kolb}.The proper time of a 4D observer (as a function of the 5th dimension) changes differently than 
the scale factor of the space coordinates. Consequently, the effective speed of light changes as a function 
of the position along the 5th dimension, and it is possible that, in a given coordinate time interval,
a gravitational wave signal traveling outside our visible universe will move over larger distances 
than light wave signals traveling inside the visible universe. 
The same effect can affect the standard model particles if they are assumed to be localized modes
of bulk fields, the particles being allowed to ``tunnel'' into the bulk (see Ref.\cite{Kraus,Dubovsky}).

In Ref.\cite{Csaki_2} the authors use the more general AdS-RN bulk and they consider only flat 3-dimensional 
spatial slices of the visible universe. Additionally they impose the existence of an event horizon in 
the bulk, and these constraints require the existence of exotic matter on the brane.

In the present work we discuss the conditions under which the geodesics will return to the brane and
we estimate the relative advance of the gravitational signal with respect to the light signal traveling 
along the brane. We will consider more generic situations, without constraining the curvature to be zero, 
but assume that the cosmic scale factor of the universe is large at present. The static and the expanding 
cases will be treated separately since we will see that we cannot take the limit $H\rightarrow 0$ in the 
expanding case. The analysis of the 5D null geodesics is essentially that of Ref.\cite{Caldwell},
(see Ref.\cite{Ishihara} for a more complete analysis including time-like and space-like geodesics) but 
in addition we explicitly impose the condition that the geodesics return to the brane. Only then such apparent
Lorentz invariance violations can be detected by a brane observer. We also use the analogy with the bending of 
light in 4D Schwarzschild spacetime, to study the bending of ``gravitational rays'' around the bulk black hole.
Since an expanding universe is not described by a Minkowski metric, therefore, 
Lorenz invariance is already broken by the expansion of the universe. We should rather call such effects 
``causality violations''. However, as explained in Ref.\cite{Kraus}, the 4D effective action for a field that can 
propagate through the bulk is not Lorentz invariant, so we call such effects Lorentz invariance violations.     

We also address the question whether such apparently superluminal signals can solve the horizon problem in 
the absence of inflation.

We find that for both the static and the expanding universes the gravitational wave traveling through
the bulk reaches a brane observer before a light wave emitted from the same source as the gravitational
wave, but traveling along the brane. Also if our brane separates two different AdSS spaces, there will 
be two different gravitational waves traveling through the two bulks, each reaching a brane observer at 
different times, both before a light wave traveling along the brane. Regarding the horizon problem, we find 
that it is possible for the bulk signals to travel far enough to homogenize the universe we see today. However,
the signals are redshifted by the gravitational field of the bulk black hole, and it is unlikely that they
will carry enough energy back to the brane to have a significant effect on homogenizing the universe.

\section{Bulk null geodesics}
\subsection{Validity of the light-ray approximation}

The light-ray approximation used to calculate the advance of the bulk gravitational waves with
respect to the electromagnetic signal traveling along the brane is valid only if the wavelength of
the graviton is much smaller than the curvature radius of the AdSS bulk\footnote{I 
thank \'E. E. Flanagan for pointing out the limits of the approximation.}: $\lambda \ll 1/l$.
If the wavelength is comparable or larger than the the AdSS curvature radius, then we should study
fluctuations around the AdSS background, describing the propagation of a gravitational wave. 
The effective 4D cosmological constant is given by:
\be
\lambda_{eff}=\frac{\lambda_{brane} ^2}4-\frac 1{l^2}
\ee
If the brane tension is of the order of $1 TeV$ then the bulk cosmological constant must be of the
same order of magnitude in order to give the small observed 4D cosmological constant. 
Consequently the calculation of the advance of the gravitational signal is valid only for gravitons with
energies of the order of $1 TeV$, impossible to observe with LIGO.

\subsection{Geodesic equations}
We study the propagation of a signal through the AdSS bulk, with the aim to find which geodesics will
leave the brane and re-intersect it. For simplicity we first study the 
stationary brane: the position of the brane along the 5th dimension simultaneously satisfies 
$H\left(r_b\right)=0$ and  $\dot H\left(r_b\right)=0$ (see appendix \ref{static_brane_solutions}). 
We use the setup of Ref.\cite{Kraus} in which the brane separates two (not necessarily identical) AdSS 
spaces described by the metric Eq.(\ref{AdDSS_metric}).
The function $f\left(r\right)$ has the general form of Ref.\cite{Birmingham}:\footnote{ As already mentioned, 
other authors \cite{Csaki_2,Csaki_3} use the more general AdS-RN space time for which the function 
$f\left(r\right)$ has one extra term: $f\left(r\right)=k+\frac{r^2}{l^2}-\frac{\mu}{r^2}+\frac{Q^2}{r^4}$}
\be
f\left(r\right)=k+\frac{r^2}{l^2}-\frac{\mu}{r^2}
\ee
We will see that the existence of Killing vectors makes (at least for the moment) the explicit form of 
$f\left(r\right)$ irrelevant.

We are interested in finding the null geodesics of the AdSS space-time. We will allow only 3 coordinates 
to change, namely, $t$, $r$ and $\sigma$. Depending on the curvature of the 3D spatial sections, this metric 
is of one of the following three forms:
\baray
&& d\Sigma_k^2=d\sigma^2+h\left(\sigma\right)\left(d\theta^2+\sin^2\theta d\phi^2\right) \nonumber \\
&& h\left(\sigma\right)=\left\{\begin{array}{lrc}
\sin^2\left(\sigma\right) & k=+1 \\
\sigma^2 & k=0 \\
\sinh^2\left(\sigma\right) & k=-1
\end{array}\right.
\earay
Since we keep $\theta$ and $\phi$ constant, the explicit form of $h\left(\sigma\right)$ is irrelevant.
With this restriction the metric for which we want to find the geodesics is:
\be
\label{short_metric}
d\hat s^2=-f\left(r\right)dt^2+r^2d\sigma^2+\frac{dr^2}{f\left(r\right)}
\ee
We notice that, up to specifying the explicit form of $f\left(r\right)$, this metric is identical with the 4D
Schwarzschild metric with the restriction $d\phi=0$. 

\begin{center}
  \epsfbox{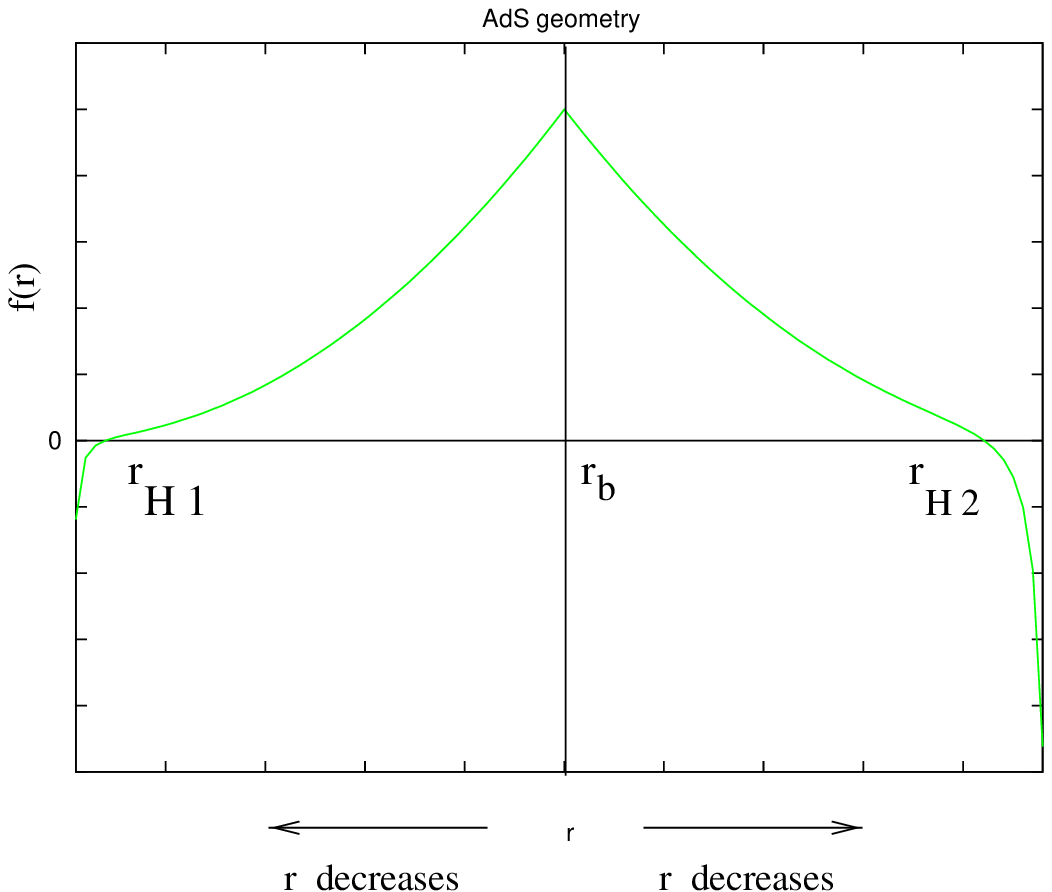}
  \parbox{12cm}{\vspace{.4cm}
    FIG.1 \hspace{2pt}The picture shows the warp factor as a function of the radial coordinate.
The brane is a domain wall placed at $r_b$, separating two (generically different) AdSS spaces
with horizons at $r_{H1}$ and $r_{H2}$, so signals can travel outside the brane through either 
space.
\vspace{12pt}}
\end{center}

We could now proceed by writing the geodesic equations,
\be
\frac{\partial^2x^{\mu}}{\partial \lambda^2}+
\Gamma_{\alpha\beta}^{\mu}\frac{\partial x^{\alpha}}{\partial \lambda}
\frac{\partial x^{\beta}}{\partial \lambda}=0
\ee
but we can use the fact that fictitious space-time described by the metric (\ref{short_metric}) possesses 
the Killing vector fields: $\left(\frac{\partial}{\partial t}\right)^{\mu}$,
$\left(\frac{\partial}{\partial \sigma}\right)^{\mu}$. We obtain two prime integrals:
\baray
&& \left(\frac{\partial}{\partial t}\right)^{\mu}\left(\frac{\partial x^{\nu}}{\partial\lambda}\right)
g_{\mu\nu}=const. \rightarrow -f\left(r\right)\frac{\partial t}{\partial \lambda}=-E \\
&& \left(\frac{\partial}{\partial \sigma}\right)^{\mu}\left(\frac{\partial x^{\nu}}{\partial\lambda}\right)
g_{\mu\nu}=const. \rightarrow r^2\frac{\partial r}{\partial \lambda}=P
\earay 
One may argue that the original metric does not possess the Killing vector field 
$\left(\frac{\partial}{\partial \sigma}\right)^{\mu}$, but the same results can be obtained by 
integrating the geodesic equations.
The geodesic equations for the $t$ and $\sigma$ coordinates are:
\baray
&& \frac{d^2t}{d\lambda^2}+\frac{1}{f\left(r\right)}\frac{\partial f\left(r\right)}{\partial r}
\frac{\partial t}{\partial\lambda}\frac{\partial r}{\partial\lambda} = 0 \\
&& \frac{d^2\sigma}{d\lambda^2}+\frac{2}{r}\frac{\partial r}{\partial\lambda}
\frac{\partial\sigma}{\partial\lambda}=0
\earay
They can be written in the form:
\baray
\frac{1}{r^2}\frac{\partial}{\partial\lambda}\left[r^2\frac{\partial\sigma}{\partial\lambda}\right]=0 
&\Longrightarrow& r^2\frac{\partial\sigma}{\partial\lambda}=const.\\
\frac{1}{f\left(r\right)}\frac{\partial}{\partial\lambda}\left[f\left(r\right)
\frac{\partial t}{\partial\lambda}\right]=0 &\Longrightarrow& 
f\left(r\right)\frac{\partial t}{\partial\lambda}=const.
\earay
which coincides with the previous result.
Using these two prime integrals and the fact that the tangent vector to the geodesic is a null vector:
\be
-f\left(r\right)\left(\frac{\partial t}{\partial \lambda}\right)^2+
r^2d\left(\frac{\partial \sigma}{\partial \lambda}\right)^2+
\frac{1}{f\left(r\right)}\left(\frac{\partial r}{\partial \lambda}\right)^2=0
\ee
we obtain the following prime integrals that allow us to discuss the properties of the geodesics.
\baray
\label{t_equation}
&& \frac{\partial t}{\partial \lambda}=\frac{E}{f\left(r\right)} \\
\label{sigma_equation}
&& \frac{\partial \sigma}{\partial \lambda}=\frac{P}{r^2} \\
\label{r_equation}
&& \frac{\partial r}{\partial \lambda}=\pm E\sqrt{1-\frac{P^2}{E^2}\frac{f\left(r\right)}{r^2}}
\earay
Using the analogy with the bending of light calculations, in FIG.2 we represent $\sigma$ as an 
angular variable.
Since in the present setup (see Ref.\cite{Kraus}) the coordinate $r$ decreases on either side of the brane,
we are only interested in geodesics that are contained in the space delimited by the brane and the horizon,
as illustrated in FIG.1.

We can now proceed to analyze the propagation of the bulk null signals and find out how they are perceived 
by a brane observer. More precisely we want to find which geodesics can start on the brane, travel through 
the bulk, and re-intersect the brane. Then we find the separation of the two events (emission, re-entry) as 
seen by a brane observer.

\begin{center}
  \epsfbox{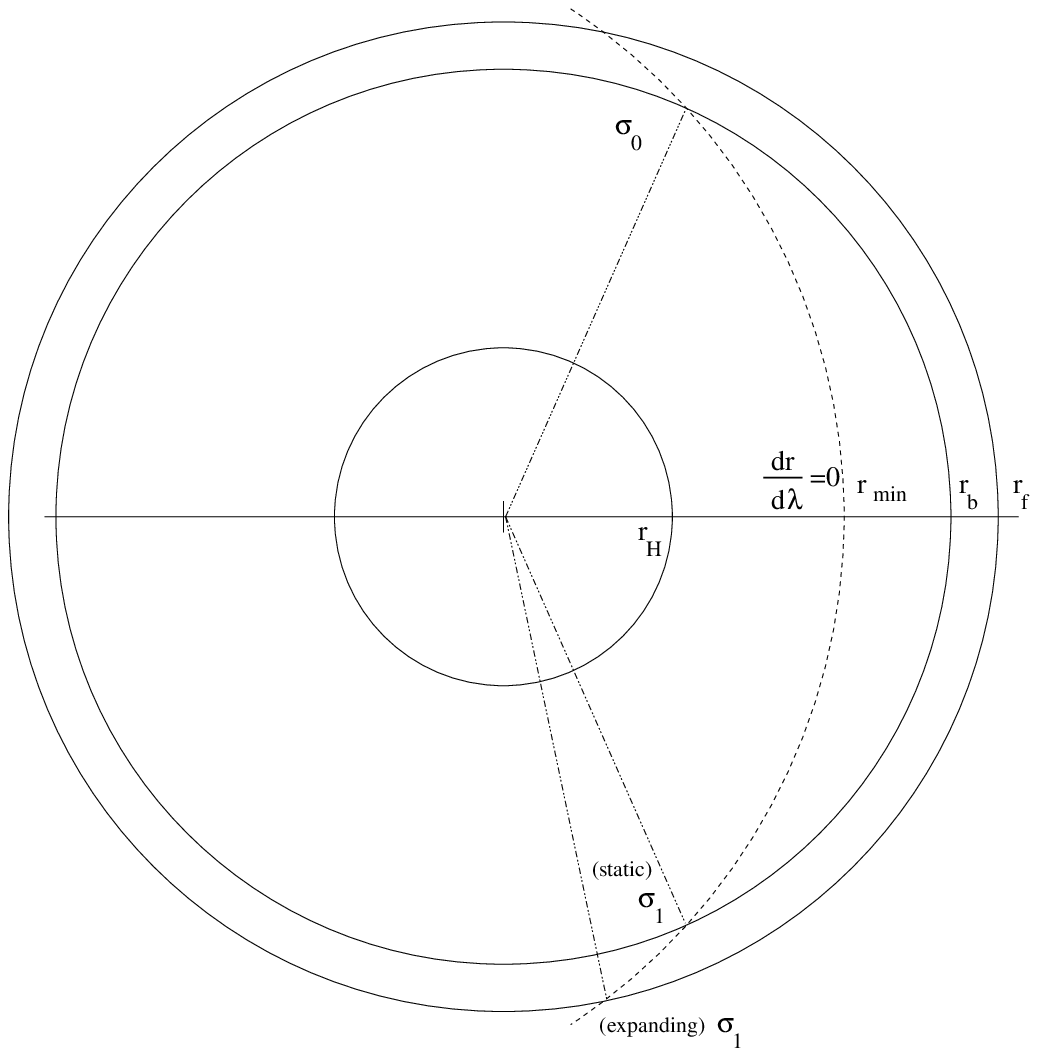}
  \parbox{12cm}{\vspace{.4cm}
    FIG.2 \hspace{2pt} The brane is located at $r=r_b$ for the static case, or moves from $r_b$ to $r_f$
in the expanding case. The geodesics that re-intersect the
brane must satisfy the condition $r_H < r_{min}$, $r_H$ being the radius of the AdSS horizon, and $r_{min}$
the location of the turning point.}
\end{center}

\section{Brane re-intersecting geodesics}
We now want to find those geodesics that will be emitted from the brane, travel through the bulk and 
re-intersect the brane. Since the $r$ coordinate decreases on both sides of the brane, the geodesics will first
move toward decreasing $r$ so in Eq.(\ref{r_equation}) we have to choose the ``-'' sign. The brane is either 
located at a fixed $r$, or moving toward larger $r$ (corresponding to an expanding universe), so in order for
the geodesics to re-intersect the brane they have to reach a turning point after which they move toward 
increasing $r$. The movement toward increasing $r$ corresponds to the ``+'' sign in  Eq.(\ref{r_equation}) so in 
order to switch between the initial ``-'' sign to the ``+'' sign, $\partial r/\partial\lambda$ must be zero at
the turning point. Moreover, the turning point must be outside the horizon of the bulk black hole, since it
will take null signals an infinite coordinate time to reach the horizon, so these signals do not re-intersect 
the brane.
 
We will study the zeroes of the function:
$F\left(r\right)=1-\frac{P^2}{E^2}\frac{f\left(r\right)}{r^2}$ since these zeroes give the location of the 
turning point of the bulk null geodesics. The equation becomes:
\be
1-\frac{P^2}{E^2}\left(\frac1{l^2}+\frac{k}{r^2}-\frac{\mu}{r^4}\right)=0
\ee
First, the locations of the AdSS horizons are given by the zeroes of the function $f\left(r\right)=
k+r^2/l^2-\mu/r^2$. The solutions are:
\be
\label{AdSS_Horizon}
\frac1{r_H^2}=\frac{k\mp\sqrt{k^2+4\frac{\mu}{l^2}}}{2\mu}
\ee
If such a horizon exists, the brane is located in the region where $f\left(r\right)>0$.
We can see the location of the zeroes  of $f\left(r\right)$ in FIG.3. Unless $\mu<0$ there is only
one horizon corresponding to the ``+'' solution (the ``-'' solution corresponds to $r_H^2<0$). 
The zeroes of $F\left(r\right)$ are given by:
\be
\frac1{r_{min}^2}=\frac{k\mp\sqrt{k^2+4\mu\left(\frac1{l^2}-\frac{E^2}{P^2}\right)}}{2\mu}
\ee
We will now discuss each signature of $\mu$ and $k$. All cases are illustrated in FIG.3.
\begin{itemize}
\item{$\mu > 0, k > 0$}
In this case there is only one horizon corresponding to the ``+'' solution. Depending on the value of $P^2/E^2$, 
$F\left(r\right)$ can have two zeros outside the horizon, in which case the geodesic corresponding to the particular 
value of $P^2/E^2$ will return at the outer one (the ``+'' solution). If the two zeroes merge we will see that the 
corresponding geodesic will not return to the brane. There will also be geodesics that do not return to the brane, because 
for small enough $P^2/E^2$, $F\left(r\right)$ will have no zeros. $F\left(r\right)$ will have no zeroes if it is positive 
at its minimum. This gives the condition: $P^2/E^2\le 1/\left(1/l^2+k^2/4\mu\right)$

\item{$\mu > 0, k < 0$}
There is one horizon corresponding to the ``+'' solution. The geodesics either have no turning point, or are not emitted 
at all. We can find values of $P^2/E^2$ such that $F\left(r\right)$ has a zero, but if $r_{min}$ satisfies the condition, 
$r_H<r_{min}<r_b$, then $F\left(r\right)$ will be negative at $r_b$, meaning that the geodesics corresponding to the values
of $P^2/E^2$ do not intersect the brane at all.

\item{$\mu < 0, k > 0$}
There is no horizon in this case, only a naked singularity at $r=0$. In this case all geodesics will
have a turning point.

\item{$\mu < 0, k < 0$}
Depending on the values of $\mu$, $k$ and $l$, there can be two horizons, one horizon, or no horizon. Let us look at the
solution Eq.(\ref{AdSS_Horizon}) and rewrite it in terms of the absolute values of $\mu$ and $k$.
\be
\frac1{r_H^2}=\frac{\left|k\right|\pm\sqrt{k^2-4\frac{\left|\mu\right|}{l^2}}}{2\mu}
\ee
Solutions exist only if $\left|\mu\right| \le k^2l^2/4$. Strict inequality corresponds to the existence of two horizons,
only the outer one (the ``+'' solution) being of interest to us.
\begin{itemize}
\item
If $\left|\mu\right| > k^2l^2/4$, there is a naked singularity at $r=0$ and all the geodesics will have a turning point.
\item
If $\left|\mu\right| = k^2l^2/4$ the two horizons merge and are located at $r_H=\sqrt{2\left|\mu\right|/\left|k\right|}$.  
If $1-\left(P^2/E^2\right)1/l^2 \ge 0$, there will be one turning point located behind the horizon, so no geodesic returns.
If $1-\left(P^2/E^2\right)1/l^2 < 0$ there will be two turning points, one outside and one inside the horizon. However, 
if we have $r_H<r_{min}$, then $F\left(r_b\right)<0$, so the corresponding geodesics do not intersect the brane at all.
\item
If  $\left|\mu\right| < k^2l^2/4$ there will be two horizons.
 $F\left(r\right)$ will have one or two zeros, depending on the value of $P^2/E^2$. For $r\rightarrow 0$,  
$F\left(r\right)$ is dominated by $-\left(P^2/E^2\right)\left(\left|\mu\right|/r^4\right)$ which is negative,
and for $r\rightarrow \infty$ by $1-\left(P^2/E^2\right)1/l^2$ which can be positive or negative. The function has 
an extremum at $r^2=2\left|\mu\right|/\left|k\right|$. The second derivative evaluated at the extremum is equal to
$-\left(P^2/E^2\right)\left(\left|k\right|^3/\left|\mu\right|^2\right)$, so the extremum is a {\em maximum}.
We thus find: If $1-\left(P^2/E^2\right)1/l^2 > 0$ the geodesics will not return to the brane since the turning
point is located behind the outer horizon. If $1-\left(P^2/E^2\right)1/l^2 < 0$, depending on the value of $P^2/E^2$,
there will be two turning points, one turning point, or no turning points depending on whether 
$1-\left(P^2/E^2\right)\left(1/l^2-k^2/4\left|\mu\right|\right)$ is positive, zero, or negative.
If there is only one turning point, then the maximum of $F\left(r\right)$ is equal to zero, so $F\left(r_b\right)<0$:
the corresponding geodesic does not intersect the brane.
If there are two turning points, one can be outside the horizon, but if $r_{min}$ of the outer turning point 
satisfies $r_H < r_{min} < r_b$, then $F\left(r_b\right)<0$ so the corresponding geodesic does not intersect the brane at all.
\end{itemize}
\end{itemize}

\begin{center}
  \epsfbox{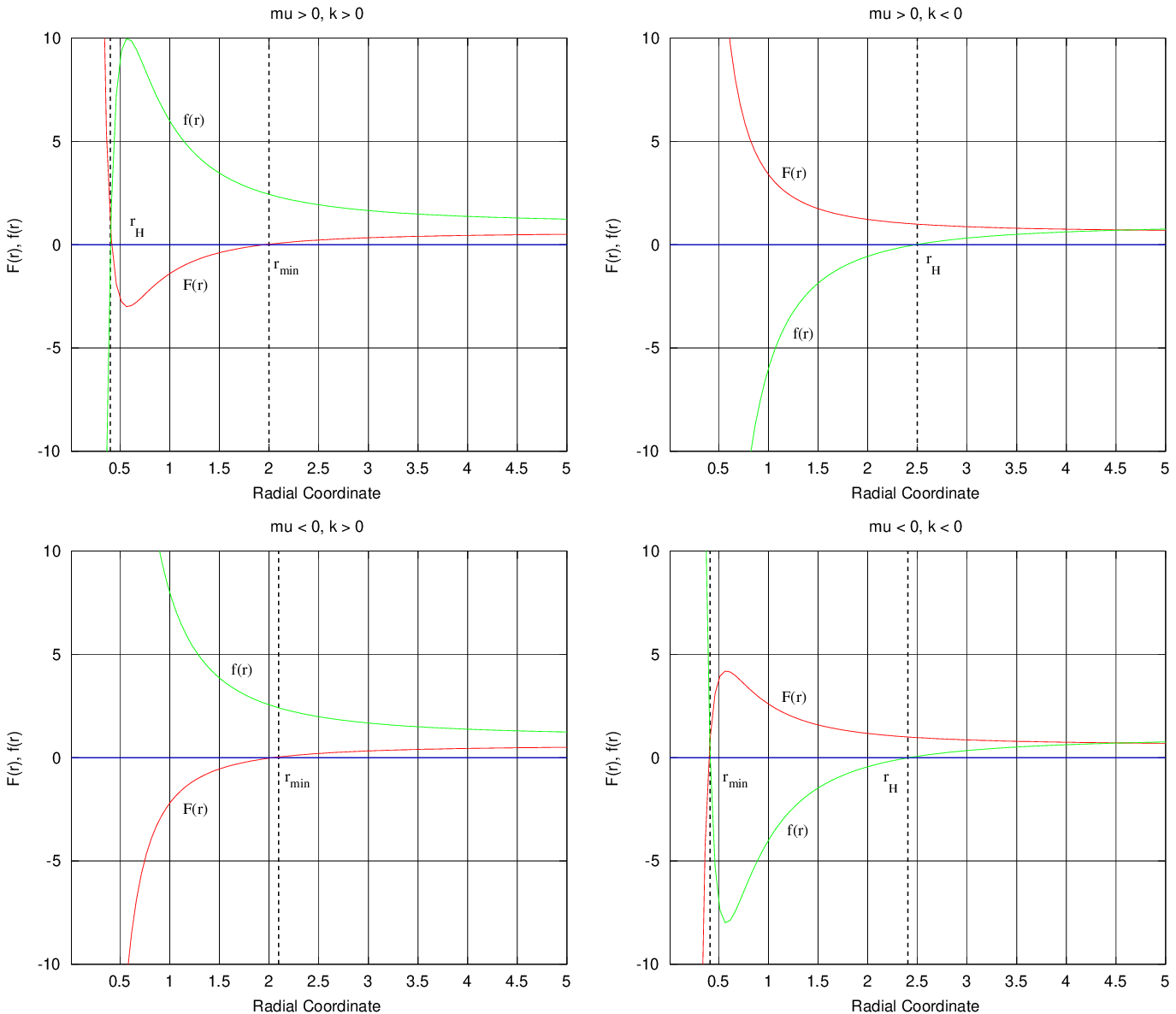}
  \parbox{12cm}{\vspace{.4cm}
   FIG.3 \hspace{2pt} The positions of the horizons are given by the zeros of $f\left(r\right)$, and the positions of 
the turning points by the zeros of $F\left(r\right)$. Their locations are marked by vertical dashed lines. The functions
$f\left(r\right)$ and $F\left(r\right)$ are plotted for all possible signatures of $\mu$ and $k$.

\vspace{12pt}}
\end{center}

\section{Static Brane}

We will now assume that the function $F\left(r\right)$ has a zero close to $r_b$, such that:
\be
\frac{\left|r_b-r_{min}\right|}{r_b} \ll1
\ee
In this case we will expand the function $F\left(r\right)$ around this zero:
\be
\label{expansion_series}
F\left(r\right)=\left[F\left(r_{min}\right)=0\right]+\left.\frac{dF}{dr}\right|_{r=r_{min}}\delta r+
\frac12\left.\frac{d^2F}{dr^2}\right|_{r=r_{min}}\left(\delta r\right)^2+O\left(\delta r^3\right)
\ee
Using the equations Eq.(\ref{t_equation}, \ref{sigma_equation}, \ref{r_equation}) we can eliminate the 
affine parameter $\lambda$ and obtain:
\be
dt=\frac{dr}{f\left(r\right)\sqrt{F\left(r\right)}}, \; d\sigma=\frac{P}{E}\frac{dr}{\sqrt{F\left(r\right)}}
\ee
Writing $r=r_{min}+\delta r$ the coordinate time and distance it takes a signal to leave and re-entry 
the brane is given by the sum of two integrals:
\be
\Delta t, \sigma = \int_{r_b}^{r_{min}}dt,\sigma+\int_{r_{min}}^{r_f}dt,\sigma
\ee
Since we chose the brane to be static, $r_f=r_b$: 
\baray
&& \Delta t=2\int_0^{r_b-r_{min}}\frac{d\left(\delta r\right)}{f\left(r_{min}+\delta r\right)
\sqrt{F\left(r_{min}+\delta r\right)}} \\
&& \Delta \sigma =2\frac{P}{E}\int_0^{r_b-r_{min}}\frac{d\left(\delta r\right)}
{\left(r_{min}+\delta r\right)^2\sqrt{F\left(r_{min}+\delta r\right)}} 
\earay
Substituting in the above equations the series expansion Eq.(\ref{expansion_series}) we observe that the 
integrals are convergent if $\left.\frac{dF}{dr}\right|_{r=r_{min}}\neq 0$, but are divergent if 
$\left.\frac{dF}{dr}\right|_{r=r_{min}}=0$. In this case it will take an infinite time for the signals to return
(see appendix \ref{divergences}).
For the case $\left.\frac{dF}{dr}\right|_{r=r_{min}}\neq 0$ we see from Eq.(\ref{r_equation}) that 
$F\left(r\right)$ is positive between $r_{min}$ and $r_b$ and since $F\left(r_{min}\right)=0$ we must have 
$\left.\frac{dF}{dr}\right|_{r=r_{min}}>0$. The function $f\left(r\right)$ has zeroes only at the location of 
the AdSS horizons, so $f\left(r_{min}\right)>0$.We can now use the expansion Eq.(\ref{expansion_series}) and 
keep only the leading order term when calculating $\Delta t$ and $\Delta \sigma$.
\baray
\label{delta_t}
&& \Delta t=2\int_0^{r_b-r_{min}}\frac{d\left(\delta r\right)}{f\left(r_{min}\right)
\sqrt{F^{\prime}\left(r_{min}\right)}\sqrt{\delta r}}=
\frac{4\sqrt{r_b-r_{min}}}{f\left(r_{min}\right)\sqrt{F^{\prime}\left(r_{min}\right)}} \\
\label{delta_sigma}
&& \Delta \sigma =2\frac{P}{E}\int_0^{r_b-r_{min}}\frac{d\left(\delta r\right)}
{\left(r_{min}\right)^2\sqrt{F^{\prime}\left(r_{min}\right)}\sqrt{\delta r}}=
\frac{4P}{E}\frac{4\sqrt{r_b-r_{min}}}{\left(r_{min}\right)^2\sqrt{F^{\prime}\left(r_{min}\right)}}
\earay
The induced metric on the brane will be:
\be
ds_b^2=-f\left(r_b\right)dt^2+r_b^2d\sigma^2
\ee
Notice that we kept the bulk coordinates to express distances on the brane, even though they are not the 
natural coordinates for the brane observer, since it will be easier to find the separation of the emission and
the re-entry. The coefficients of the metric are constant, consequently the separation of the emission and re-entry 
will be:
\be
\Delta s^2=-f\left(r_b\right)\Delta t^2+r_b^2\Delta \sigma^2
\ee
Using the above results for $\Delta t$ and $\Delta \sigma$ we want to find whether $\Delta s^2>0$ or 
$\Delta s^2<0$. For calculational convenience we will work with the ratio $\Delta s^2/r_b^2\Delta \sigma^2$.
\baray
&& \frac{\Delta s^2}{r_b^2\Delta \sigma^2}=1-\frac{f\left(r_b\right)}{r_b^2}
\left(\frac{\Delta t}{\Delta \sigma}\right)^2 = 1-\frac{f\left(r_b\right)}{r_b^2}
\frac{\left[\frac1{f\left(r_{min}\right)\sqrt{F^{\prime}\left(r_{min}\right)}}\right]^2}
{\left[\frac{P}{E}\frac1{\left(r_{min}\right)^2\sqrt{F^{\prime}\left(r_{min}\right)}}\right]^2} \nonumber \\
&& =1-\frac{f\left(r_b\right)}{r_b^2}\frac{E^2}{P^2}\frac{r_{min}^4}{f^2\left(r_{min}\right)}
\earay
We use the fact that 
\be
F\left(r_{min}\right)=1-\frac{P^2}{E^2}\frac{f\left(r_{min}\right)}{r_{min}^2}=0 \; \Longrightarrow \;
\frac{r_{min}^2}{f\left(r_{min}\right)}=\frac{P^2}{E^2} 
\ee
to obtain:
\be
\label{relative+advance}
\frac{\Delta s^2}{r_b^2\Delta \sigma^2}=1-\frac{f\left(r_b\right)}{r_b^2}\frac{E^2}{P^2}\frac{P^4}{E^4}=
1- \frac{f\left(r_b\right)}{r_b^2}\frac{P^2}{E^2}=F\left(r_b\right)>0
\ee
We obtained that $\Delta s^2>0$, consequently the brane observer will see the two events as spatially 
separated, so the bulk signal travels faster than light on the brane.

We now try to express the ratio $\frac{\Delta s^2}{r_b^2\Delta \sigma^2}$ in terms of the scale factor of
the visible universe, $r_b$, and the physical distance separating the two events as seen by a brane observer. 
Using the solution (\ref{delta_sigma}), and expressing $P^2/E^2$ in terms of
$r_{min}$, we obtain:
\be
\label{r_min_only_sigma}
\Delta\sigma = \frac{4\sqrt{r_b-r_{min}}}{r_{min}\sqrt{\frac{2f\left(r_{min}\right)}{r_{min}}
-f^{\prime}\left(r_{min}\right)}}
\ee

We see that we can now express $r_{min}$ in terms of $\Delta\sigma$ and $r_b$, so we can express the relative advance
of the gravitational signal in terms of the physical separation of the emission and re-entry points on the brane,
$r_b\Delta\sigma$, and the parameters of the AdSS bulk, namely, $k$, $l$ and $\mu$.
Using Eq.(\ref{relative+advance}) and the fact that $F\left(r_{min}\right)=0$, we obtain:
\be
\frac{\Delta s^2}{r_b^2\Delta \sigma^2}=F^{\prime}\left(r_{min}\right)
\left(r_b-r_{min}\right)+O\left(r_b-r_{min}\right)^2
\ee
Now we can further use the fact that $r_b-r_{min}$ is a small quantity and write:
\be
F^{\prime}\left(r_{min}\right)=F^{\prime}\left(r_b\right)+F^{\prime\prime}\left(r_b\right)
\left(r_b-r_{min}\right)+O\left(r_b-r_{min}\right)^2
\ee
so we obtain (up to corrections of order $O\left(r_b-r_{min}\right)^2$) 
\be
\frac{\Delta s^2}{r_b^2\Delta \sigma^2}\approx F^{\prime}\left(r_b\right)\left(r_b-r_{min}\right)\approx
\left(\frac{2}{r_b}-\frac{f^{\prime}\left(r_b\right)}{f\left(r_b\right)}\right)\left(r_b-r_{min}\right)
\ee
We now use Eq.(\ref{r_min_only_sigma}) and again the fact that $r_b-r_{min}$ is a small quantity:
\be
\Delta\sigma = \frac{4\sqrt{r_b-r_{min}}}{r_{min}\sqrt{f\left(r_{min}\right)}
\sqrt{\frac{2}{r_{min}}-\frac{f^{\prime}\left(r_{min}\right)}{f\left(r_{min}\right)}}}\approx
\frac{4\sqrt{r_b-r_{min}}}{r_b\sqrt{f\left(r_b\right)}
\sqrt{\frac{2}{r_b}-\frac{f^{\prime}\left(r_b\right)}{f\left(r_b\right)}}}
\ee
We obtain:
\be
r_b-r_{min}=\left(\frac{\Delta\sigma}{4}\right)^2 r_b^2 f\left(r_b\right)
\left(\frac{2}{r_b}-\frac{f^{\prime}\left(r_b\right)}{f\left(r_b\right)}\right)
\ee
and the final form for the relative advance of the gravitational signal with respect to the light signal 
along the brane:
\be
\frac{\Delta s^2}{r_b^2\Delta \sigma^2}\approx\frac1{16}f\left(r_b\right)
\left(\frac{2}{r_b}-\frac{f^{\prime}\left(r_b\right)}{f\left(r_b\right)}\right)^2\left(\Delta L\right)^2
\ee
where $\Delta L=r_b\Delta\sigma$ is the physical separation between the emission and re-entry points of 
the gravitational signal.

\section{Expanding brane}
We will now consider the case where the 4D Hubble constant is non-zero. We obtain a non-zero 4D Hubble 
constant by allowing the brane to move along the 5th dimension, $r_b=r_b\left(t\right)$. We want to find 
the proper time of an observer moving with the brane. We therefore choose an observer whose world-line is
described by $\sigma, \theta, \phi=const.$. Following Ref.\cite{Caldwell} we express the proper time of the
brane in terms of the $5D$ coordinate time. 
\be
\label{brane_proper_time}
ds_{brane}^2=-d\tau^2=-f\left(r_b\left(t\right)\right)dt^2+\frac{dr_b^2}{f\left(r_b\left(t\right)\right)}=
-f\left(r_b\left(t\right)\right)dt^2+\frac{\dot r_b^2d\tau^2}{f\left(r_b\left(t\right)\right)}
\ee
where the dot represents the derivative taken with respect to the proper time of the brane $\tau$. 
The induced metric on the brane becomes:
\be
ds_{brane}^2=-d\tau^2+r_b^2\left(\tau\right)d\Sigma_k^2
\ee
so the Hubble constant seen by a brane observer is: $H=\dot r_b/r_b$. We can now use
Eq.(\ref{brane_proper_time}) to express $dr_b/dt$ as a function of $H=\dot r_b/r_b$. 
\be
f\left(r_b\left(t\right)\right)dt^2=d\tau^2\left(1+\frac{\dot r_b^2}{f\left(r_b\left(t\right)\right)}\right)
=d\tau^2\left(1+\frac{H^2r_b^2}{f\left(r_b\left(t\right)\right)}\right)
\ee
\be
\frac{dr_b}{dt}=\frac{dr_b}{d\tau}\frac{d\tau}{dt}=
\dot r_b\sqrt{\frac{f\left(r_b\right)}{1+\frac{H^2r_b^2}{f\left(r_b\right)}}}=
Hr_b\sqrt{\frac{f\left(r_b\right)}{1+\frac{H^2r_b^2}{f\left(r_b\right)}}}
=\frac{Hr_bf\left(r_b\right)}{\sqrt{f\left(r_b\right)+H^2r_b^2}}
\ee
We can now compare the movement of the brane with the movement of bulk null signals so that we can 
identify the signals that will re-entry the brane. For these signals, since the 
4D metric is no longer time-independent, we will calculate the coordinate distance traveled by a 
null signal along the brane, and compare it with the coordinate distance along the brane traveled by the 
bulk signal.
\be
\Delta\sigma_{brane \; null}=\int_{r_{initial}}^{r_{final}}\frac{d\tau}{r_b\left(\tau\right)}=
\int_{r_{initial}}^{r_{final}}\frac{dr_b}{Hr_b^2}
\ee
We will assume that the Hubble constant is (approximately) a true constant, namely that $r_b$ is 
sufficiently large that between the emission and re-entry of the bulk signal the amount of brane expansion 
satisfies the condition:
\be
\frac{\Delta r_b}{r_b} \ll 1
\ee
Also for large $r_b$ the function $f\left(r\right)=k+\frac{r^2}{l^2}-\frac{\mu}{r^2}$ is dominated by the 
term proportional to $r^2$, so that for the motion of the brane we obtain the equation:
\be
\label{brane_movement}
\frac{dr_b}{dt}=\frac{Hr_bf\left(r_b\right)}{\sqrt{f\left(r_b\right)+H^2r_b^2}}\approx
\frac{\frac{Hr_b^3}{l^2}}{\sqrt{\left(H^2+\frac1{l^2}\right)r_b^2}}=
\frac{Hr_b^2}{l\sqrt{1+H^2l^2}}
\ee
The resulting equation for the evolution of the position of the brane is easy to integrate:
\be
\label{exp_r_brane}
\frac{dr_b}{r_b^2}=\frac{H}{l}\frac{dt}{\sqrt{1+H^2l^2}}\;\Longrightarrow\;
\frac1{r_0}-\frac1{r_f}=\frac{H}{l}\frac{t}{\sqrt{1+H^2l^2}}
\ee
The equation of motion for null signals through the bulk is given by:
\be
\label{signal_movement}
\frac{dr}{dt}=\pm f\left(r\right)\sqrt{F\left(r\right)}
\ee
Initially $r$ will decrease to $r_{min}$ $(F\left(r_{min}\right)=0)$, so we will have to use the ``-'' 
solution. After reaching $r_{min}$, $r$ will increase, so we will have to consider the ``+'' solution.
We will continue to use the the expansion around $r_{min}$, so we can use the previous result 
Eq.(\ref{delta_t}):
\be
\label{exp_r_signal}
\Delta t=\frac{2\sqrt{r_0-r_{min}}}{f\left(r_{min}\right)\sqrt{F^{\prime}\left(r_{min}\right)}}+
\frac{2\sqrt{r_f-r_{min}}}{f\left(r_{min}\right)\sqrt{F^{\prime}\left(r_{min}\right)}}
\ee
where $r_f$ is the $r$-coordinate where the bulk signal re-enters the brane. The time coordinate for the 
re-entry of the bulk signal is the intersection of the trajectories described by the equations 
Eq.(\ref{brane_movement}) and Eq.(\ref{signal_movement}), having the initial conditions 
$r_b\left(0\right)=r_0$ and $r\left(0\right)=r_0$. So we will set $r_f$ to be the same in both 
Eq.(\ref{exp_r_brane}) and Eq.(\ref{exp_r_signal}) and solve for $t$. We use the fact that the relative 
increase in $r_b$ is small to linearize Eq.(\ref{brane_movement}).
\be
\frac1{r_0}-\frac1{r_f}\approx\frac{r_f-r_0}{r_0^2}=\frac{H}{l}\frac{\Delta t}{\sqrt{1+H^2l^2}}
\ee
At the same time we rewrite Eq.(\ref{exp_r_signal}) in the form:
\be
\sqrt{r_f-r_{min}}=
\frac{f\left(r_{min}\right)\sqrt{F^{\prime}\left(r_{min}\right)}\Delta t}2-\sqrt{r_0-r_{min}}
\ee
so we obtain:
\baray
&& r_f-r_0=\left(\frac{f\left(r_{min}\right)\sqrt{F^{\prime}\left(r_{min}\right)}\Delta t}2-
\sqrt{r_0-r_{min}}\right)^2-\left(r_0-r_{min}\right)= \nonumber \\
&& \frac{f\left(r_{min}\right)\sqrt{F^{\prime}\left(r_{min}\right)}\Delta t}2
\left(\frac{f\left(r_{min}\right)\sqrt{F^{\prime}\left(r_{min}\right)}\Delta t}2-
2\sqrt{r_0-r_{min}}\right)=r_0^2\frac{H}{l}\frac{\Delta t}{\sqrt{1+H^2l^2}}
\earay
so we obtain the re-entry time for the bulk signal:
\be
\Delta t=\frac{4\sqrt{r_0-r_{min}}}{f\left(r_{min}\right)\sqrt{F^{\prime}\left(r_{min}\right)}}+
\frac{4r_0^2H}{f^2\left(r_{min}\right)F^{\prime}\left(r_{min}\right)l\sqrt{1+H^2l^2}}
\ee
The solution can be seen in the picture below:
\begin{center}
  \epsfbox{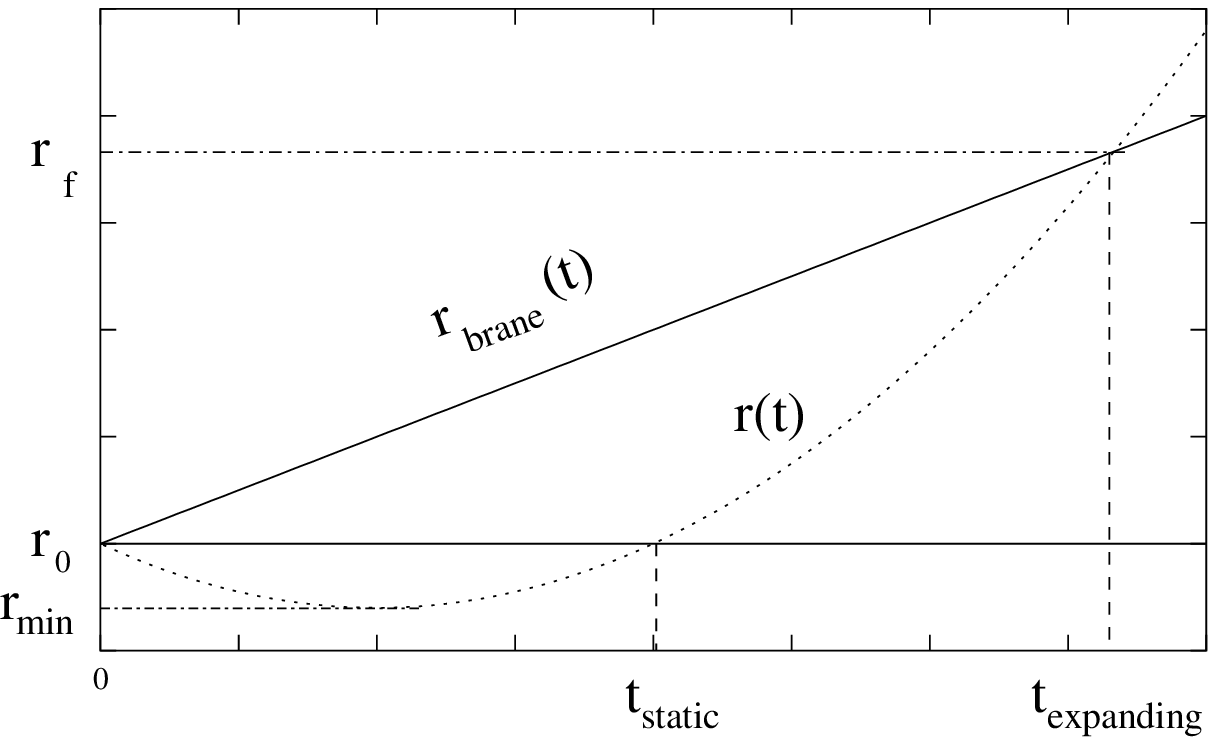}
  \parbox{12cm}{\vspace{.4cm}
    FIG.4 \hspace{2pt} The intersection of the two trajectories $r\left(t\right)$ and $r_{brane}\left(t\right)$
is the re-entry time for the bulk signal. We observe that for $H=0$ we reproduce the result from the static case. 
$t_{static}$ and $t_{expanding}$ are the re-entry times for the static and the expanding cases. \vspace{12pt}}
\end{center}

We now go back and calculate $\Delta\sigma$ for the brane signal and for the bulk signal. Again, we assume
that $H\approx const.$ and that $r$ is large so that the relative expansion of the brane is small, 
$\Delta r_b/r_b \ll 1$, and $f\left(r\right)\approx r^2/l^2$. For the brane null signal we have:
\be
\Delta\sigma_{brane \; null}=\int_{r_0}^{r_f}\frac{dr_b}{Hr_b^2}\approx\frac1H\left(\frac1{r_0}-\frac1{r_f}\right)
\ee
We use now Eq.(\ref{exp_r_brane}) to obtain:
\be
\label{sigma_brane}
\Delta\sigma_{brane \; null}=\int_{r_0}^{r_f}\frac{dr_b}{Hr_b^2}\approx\frac1H\left(\frac1{r_0}-\frac1{r_f}\right)=
\frac1H\frac{H}{l}\frac{\Delta t}{\sqrt{1+H^2l^2}}=\frac1{\sqrt{1+H^2l^2}}\frac1l \Delta t
\ee
For the bulk null signal we have:
\baray
&& \Delta\sigma_{bulk \; null}=-\frac{P}{E}\int_{r_0}^{r_{min}}\frac{d\left(\delta r\right)}{\left(r_{min}\right)^2
\sqrt{F^{\prime}\left(r_{min}\right)}\sqrt{\delta r}}+
\frac{P}{E}\int_{r_{min}}^{r_f}\frac{d\left(\delta r\right)}{\left(r_{min}\right)^2
\sqrt{F^{\prime}\left(r_{min}\right)}\sqrt{\delta r}} \nonumber \\
&& \frac{2P}{E}\frac1{\left(r_{min}\right)^2\sqrt{F^{\prime}\left(r_{min}\right)}}\left(
\sqrt{r_0-r_{min}}+\sqrt{r_f-r_{min}}\right)=\frac{P}{E}\frac{f\left(r_{min}\right)}{\left(r_{min}\right)^2}
\Delta t
\earay
Using the fact that $P^2/E^2=\left(r_{min}\right)^2/f\left(r_{min}\right)$, 
and the approximation $f\left(r\right)\approx r^2/l^2$ we obtain:
\be
\Delta\sigma_{bulk \; null}=\frac1l \Delta t
\ee
Comparing this result with the one from Eq.(\ref{sigma_brane}) we obtain immediately that:
\be
\Delta\sigma_{brane \; null} < \Delta\sigma_{bulk \; null}
\ee
so the bulk signal will appear as superluminal to the brane observer. 
We see now why we should consider the static and expanding brane cases separately; taking the limit 
$H\rightarrow 0$ in  Eq.(\ref{sigma_brane}) would give 
$\Delta\sigma_{brane \; null} =\Delta\sigma_{bulk \; null}$, but the limit is meaningless since 
$\Delta\sigma_{brane \; null}$ given by Eq.(\ref{sigma_brane}) naively diverges. We should take into account
that as $H\rightarrow 0$ then $r_f\rightarrow r_0$ and the combination $1/H\left(1/r_0-1/r_f\right)$ is finite.

\section{The horizon problem}

We want to address the question whether the apparently superluminal signals can solve the horizon problem in the 
absence of inflation.
Inflationary models assume that the universe starts in a radiation-dominated phase during which it can become
homogeneous, followed by the inflationary phase, when the comoving horizon, defined as $1/HR\left(\tau\right)$, 
shrinks by a factor of at least $e^{55}$. 
After the end of inflation, the comoving horizon increases again, but regions that were causally 
disconnected during inflation look identical (on sufficiently large scales) because they were in causal contact
before the inflationary epoch.

In order for the bulk signals to homogenize the universe on comoving scales that become visible today,
the signals should bridge distances much larger than the size of the horizon at the moment of emission. Moreover,
we expect such signals to be emitted when the temperature was larger than $\sim 1TeV$ and re-enter the
visible universe before the recombination epoch. The reason is that once structure (galaxy clusters) forms, such 
signals should have little effect on homogenizing the universe, as the re-entry of a bulk signal should manifest itself
as a flash of energy coming from the decay of an invisible particle. It is hard to imagine how such processes 
can generate a homogeneous universe without wiping out any structure already formed. Therefore we consider only 
signals emitted when $T\sim 1TeV$ and re-enter the universe the latest when $T\sim 0.2eV$, corresponding to
the recombination epoch.

However, when the temperature is  $T\sim 1TeV$ the universe is radiation-dominated and expands rapidly, and in
our model the expansion is described by the movement of the brane away from the bulk black hole.
The signals are emitted from the brane close to the bulk black hole and will re-enter the brane farther away 
from it, so they are red-shifted in the gravitational field of the bulk black hole. Moreover
the signals are emitted from a moving brane, so we should take into account the additional Doppler shift. 

We now address the question whether it is possible for bulk signals to travel far enough to homogenize the 
visible universe. In order to do so we analyze the travel of a signal moving at the speed of light along the 
brane. The signal was emitted when the temperature was $\sim 1TeV$, and we want to find the comoving distance it
travels until the present epoch. If there is no inflation, there are two relevant parts of the journey: In the first
phase the universe is radiation-dominated, and after matter-radiation decoupling the universe becomes 
matter-dominated. For the radiation-dominated universe the energy density decreases with the expansion like
$\rho \sim 1/R^4$ while for the matter-dominated era $\rho \sim 1/R^3$. 

For a null signal traveling along the brane have:
\be
ds^2=-d\tau^2+R^2\left(\tau\right)d\sigma^2=0 \Longrightarrow
d\sigma = \frac{d\tau}{R\left(\tau\right)} = \frac{dR}{HR^2\left(\tau\right)}
\ee
The Hubble constant is given by:
\be
H^2=\frac{8\pi G}{3}\rho = \left\{\begin{array}{ccc}
\sim 1/R^4 & , & radiation \; dominated \; universe \\
\sim 1/R^3 & , & matter \; dominated \; universe \\
\end{array}\right.
\ee
so that we obtain the comoving distance traveled in terms of the scale factor of the universe (or redshift):
\be
\sigma-\sigma_0 = \left\{\begin{array}{ccc} 
\frac{R-R_0}{\sqrt{8\pi G\rho_0R_0^4}} & , & radiation \; dominated \; universe \\
\frac{\sqrt{R}-\sqrt{R_0}}{\sqrt{8\pi G\rho_0R_0^3}} & , & matter \; dominated \; universe \\
\end{array}\right.
\ee

The decoupling of radiation from matter occurs at a redshift of $10^3$, while the emission of the signal
occurs at a redshift of $10^{16}$, corresponding to a temperature of $1TeV$. Therefore most of the comoving distance 
traveled by the null signal along the brane comes from the matter-dominated era. At the same time, the bulk signal
must travel at least the same comoving distance and re-enter the brane before decoupling, namely at a redshift 
$>10^3$.
The equation describing the movement of the bulk signal is:
\be
d\sigma = \frac{P}{E}\frac{dr}{\sqrt{F\left(r\right)}}
\ee
and for large values of $R$ the comoving distance traveled by the bulk signal is: 
\be
\label{result_bulk}
\sigma - \sigma_0 = \frac{R-R_0}{\sqrt{\frac{E^2}{P^2}-\frac{1}{l^2}}}
\ee
We see that for the radiation dominated universe both comoving distances depend linearly on the scale factor, while 
for the matter-dominated universe the bulk signal can travel ``faster'' than the signal traveling along the brane. 
We can always choose $E/P$ such that the denominator of Eq.(\ref{result_bulk}) is small and the comoving distance 
traveled by the bulk signal exceeds the one traveled by the brane signal by a large factor. 
It is therefore possible for the bulk signal to travel the comoving distance required to homogenize the visible 
universe. However, we must still address the question whether the signal will deposit a sizable amount of energy on
the brane upon re-entry.
\begin{center}
\epsfbox{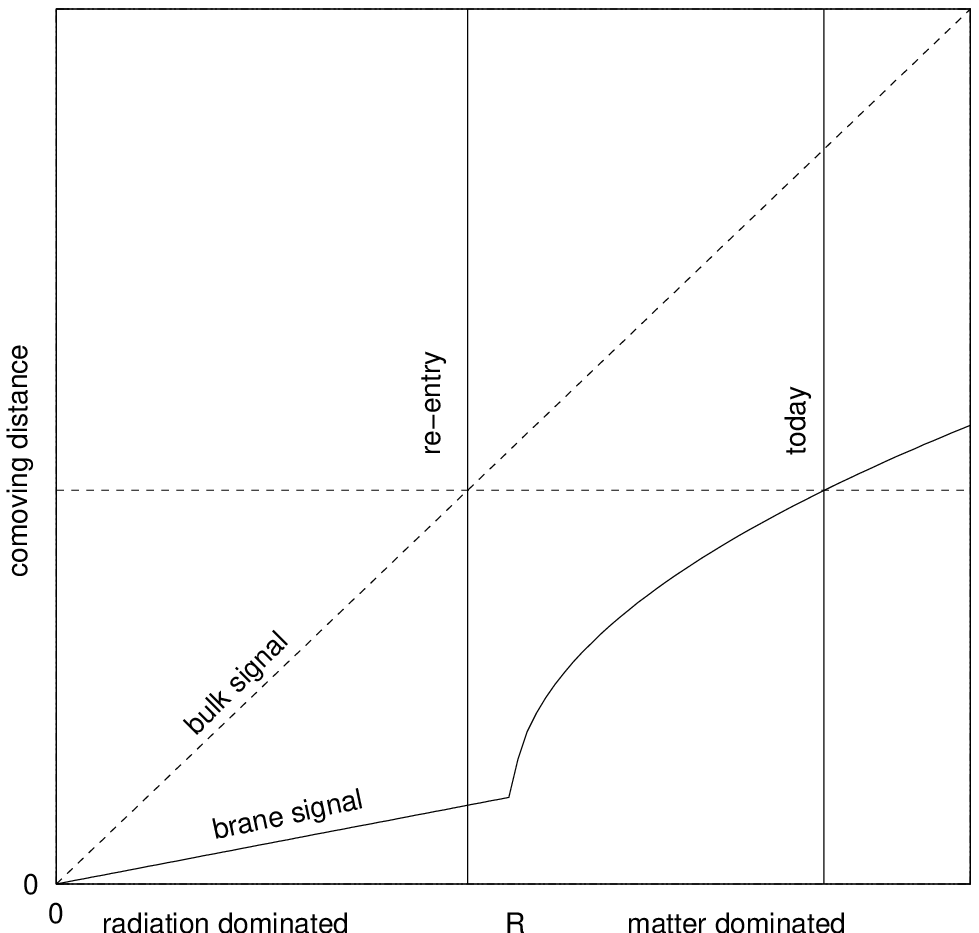}
\parbox{12cm}{\vspace{6pt} FIG.5 \hspace{2pt} Comoving distance traveled by the bulk and the brane null signals 
as a function of the scale factor of the universe (redshift). We see that at the moment of re-entry the bulk 
signal has traveled a much larger comoving distance compared with the brane signal.}
\end{center}

\subsection{Red-shift of the bulk signal}

In order to study the emission and re-entry of the signal we use the equation of motion of the brane 
and the equation of motion of the bulk signal:
\be
\frac{dr_b}{dt}=\frac{Hr_bf\left(r_b\right)}{\sqrt{f\left(r_b\right)+H^2r_b^2}}\;,\; 
\frac{dr}{dt}=\pm f\left(r\right)\sqrt{F\left(r\right)}
\ee

The signal is emitted when the brane is located at $r_{em}$ and re-enters (it is absorbed) when the brane 
is located at $r_{ab}$. We integrate the above equations:
\baray
\label{time_brane}
t_{ab}-t_{em} &=& \int_{r_{em}}^{r_{ab}}\frac{\sqrt{f\left(r_b\right)+H^2r_b^2}}{Hr_bf\left(r_b\right)}dr_b \\
\label{time_bulk}
t_{ab}-t_{em} &=& \int_{r_{em}}^{r_{min}}-\frac{dr}{f\left(r\right)\sqrt{F\left(r\right)}}+
\int_{r_{min}}^{r_{ab}}\frac{dr}{f\left(r\right)\sqrt{F\left(r\right)}}
\earay
We now consider two signals emitted along the same trajectory at a short time interval one after the other, 
more precisely, at moments $t_{em}$ and $t_{em} + \delta t_{em}$ . We want to find the time interval between the 
re-entry of the two signals, so we denote the re-entry moments by  $t_{ab}$ and $t_{ab} + \delta t_{ab}$. 
We can  use any one of Eq.(\ref{time_brane},\ref{time_bulk}), but we prefer Eq.(\ref{time_bulk}) because 
the Doppler effect mentioned before is readily visible.
\be
\label{shifted_time_bulk}
\left(t_{ab} + \delta t_{ab}\right) - \left(t_{em} + \delta t_{em}\right) = 
\int_{r_{em}+\delta r_{em}}^{r_{min}}-\frac{dr}{f\left(r\right)\sqrt{F\left(r\right)}}+
\int_{r_{min}}^{r_{ab}+\delta r_{ab}}\frac{dr}{f\left(r\right)\sqrt{F\left(r\right)}}
\ee
Subtracting Eq.(\ref{time_bulk}) from Eq.(\ref{shifted_time_bulk}), we obtain:
\be
\label{diff_coord_time}
\delta t_{ab} - \delta t_{em} = \frac{\delta r_{em}}{f\left(r_{em}\right)\sqrt{F\left(r_{em}\right)}} + 
\frac{\delta r_{ab}}{f\left(r_{ab}\right)\sqrt{F\left(r_{ab}\right)}}
\ee
We can understand the meaning of the above equation by looking at FIG.(2). The induced metric on the brane 
is $ds^2=-d\tau^2+R^2\left(\tau\right)d\sigma^2$ and since the brane expands, in the $5D$ description it moves
towards increasing $R$, i.e. away from the black hole.
The signal is first emitted towards the bulk black hole, and in the direction opposite to the 
movement of the brane. This results in the first red-shift term in Eq.(\ref{shifted_time_bulk}). When the 
signal re-enters the brane, both the brane and the signal move towards increasing $R$, which results
in the second red-shift term. 

We now want to relate $\delta t_{ab}$, $\delta t_{em}$ and $\delta r_{ab}$, $\delta r_{em}$ with the proper 
time of a brane observer. By doing so we get an estimate of the red-shift of the signal as seen by a brane
observer. The coordinate time, $t$, and the proper time of a brane observer, $\tau$, are related by:
\be
dt = d\tau\sqrt{\frac{1}{f\left(r_b\right)}\left(1+\frac{H^2r_b^2}{f\left(r_b\right)}\right)}
\ee
We can use the definition of the Hubble constant to relate $\delta r_{ab}$, $\delta r_{em}$, with the proper 
time of a brane observer. 
\be
H = \frac1R\frac{dR}{d\tau} \Longrightarrow \delta r_{em} = H_{em}r_{em}\delta\tau_{em}\;,\;
\delta r_{ab} = H_{ab}r_{ab}\delta\tau_{ab}
\ee
We now use these results in Eq.(\ref{diff_coord_time}) and obtain:
\be
\frac{d\tau_{ab}}{d\tau_{em}}=
\frac{\sqrt{\frac{1}{f\left(r_{em}\right)}\left(1+\frac{H_{em}^2r_{em}^2}{f\left(r_{em}\right)}\right)}+
\frac{H_{em}r_{em}}{f\left(r_{em}\right)\sqrt{F\left(r_{em}\right)}}}
{\sqrt{\frac{1}{f\left(r_{ab}\right)}\left(1+\frac{H_{ab}^2r_{ab}^2}{f\left(r_{ab}\right)}\right)}-
\frac{H_{ab}r_{ab}}{f\left(r_{ab}\right)\sqrt{F\left(r_{ab}\right)}}}
\ee
We see that in the absence of expansion ($H=0$), we obtain the usual result for the red-shift of a signal in the 
gravitational field of a black hole: 
\be
\frac{d\tau_{ab}}{d\tau_{em}}=\sqrt{\frac{f\left(r_{ab}\right)}{f\left(r_{em}\right)}}
\ee
Assuming that both $r_{em}$ and $r_{ab}$ are large compared with the radius of the bulk black hole such that 
$f\left(r\right)\sim r^2/l^2$, and that the expansion rate is small compared to the radial speed of light we obtain:
\be
\frac{d\tau_{ab}}{d\tau_{em}}\sim\frac{r_{ab}}{r_{em}}
\ee
namely, the bulk signal is red-shifted exactly like radiation on the brane. \footnote{We would expect this result if we
use the holographic description for the brane-world. The bulk signals are seen as excitations of some brane CFT, and 
the energy of the excitations red-shifts the same way as radiation on the brane.} 

The typical energy of the bulk signals at emission is the temperature of the radiation on the brane, so at the 
moment of re-entry the signal must have an energy equal to the new temperature of the radiation. The Doppler 
effect only increases the redshift of the bulk signal. The energy of the signal re-entering the brane being smaller 
than the typical energy of a CMB photon, we will need large numbers of bulk signals, comparable to the number
of photons in the CMB, in order to homogenize the universe. In such a situation a dangerously large amount of energy 
will leak into the bulk, possibly changing the result of nucleosynthesis. We therefore expect that the only 
contribution of the bulk signals is a deformation of the black-body spectrum of the Cosmic Microwave Background.

\section{Acknowledgments}

I thank Henry Tye, Ira Wasserman, \'Eanna Flanagan, Csaba Csaki, Nick Jones, K.Narayan, Etienne Racine, 
Saswat Sarangi and Anupam Mazumdar for discussions.
This research is partially supported by NSF.

\appendix

\section{Static brane solutions}
\label{static_brane_solutions}
We now look for solutions of the equations $H=0$ and $\dot H=0$. We use the general solution for $H$ found 
in Ref.\cite{io} (See also Ref.\cite{Gregory}). The relevant equation is:
\baray
\label{H_Kraus}
&& H^2=\frac{\lambda ^2}4-\frac 12\left( \frac 1{l_-^2}+\frac 1{l_+^2}\right) +
\frac 1{4\lambda ^2}\left( \frac 1{l_-^2}-\frac 1{l_+^2}\right) ^2+ 
\frac {k}{r^2} \nonumber \\
&& +\frac 1{r^4}\left\{ \frac{\mu_-+\mu_+}2-\frac{\mu_+-\mu_-}{
2\lambda ^2}\left( \frac 1{l_-^2}-\frac 1{l_+^2}\right) \right\} +
\frac 1{r^8}\frac{\left( \mu_+-\mu_-\right) ^2}{4\lambda ^2}
\earay
For the symmetric case $l_+=l_-$, $\mu_+=\mu_-$, the equation simplifies to:
\be
H^2=\frac{\lambda ^2}4-\frac 1{l^2}+\frac{k}{r^2}+\frac{\mu}{r^4}
\ee
In this case the conditions $H=0$ and $\dot H=0$ will give the constraints:
\be
r_b^2=-\frac{k}{2\mu} \hspace{12pt} and \hspace{12pt} \frac{\lambda ^2}4-\frac 1{l^2}-\frac{k^2}{4\mu}=0
\ee
Since we want to avoid the presence of naked singularities in the bulk, we consider only solutions 
with $\mu>0$, which will restrict us to $k<0$. In order to avoid this constraint we will allow for
$\mu_+\neq\mu_-$, while keeping $l_+=l_-$ for calculational convenience. 
In this case there will be two "superluminal" bulk signals, one through each AdSS space, arriving at the brane
at different times since $\mu_+\neq\mu_-$. The dependence of the Hubble constant on the scale factor of the 
visible universe:
\be
H^2=\frac{\lambda ^2}4-\frac 1{l^2}+\frac{k}{r^2}+\frac{\mu_++\mu_-}{2r^4}+
\frac1{r^8}\frac{\left( \mu_+-\mu_-\right)^2}{4\lambda^2}
\ee
is given in FIG. 5. We see that with the exception of the case 
$\mu_{avg}=\left(\mu_++\mu_-\right)/2<0$, $k>0$, when there is no extremum for $H\left(r\right)$, 
we can adjust the parameters $\lambda$, $k$, $l$, $\mu_+$,$\mu_-$, so that the extremum occurs for $H\left(r\right)=0$.

\begin{center}
  \epsfbox{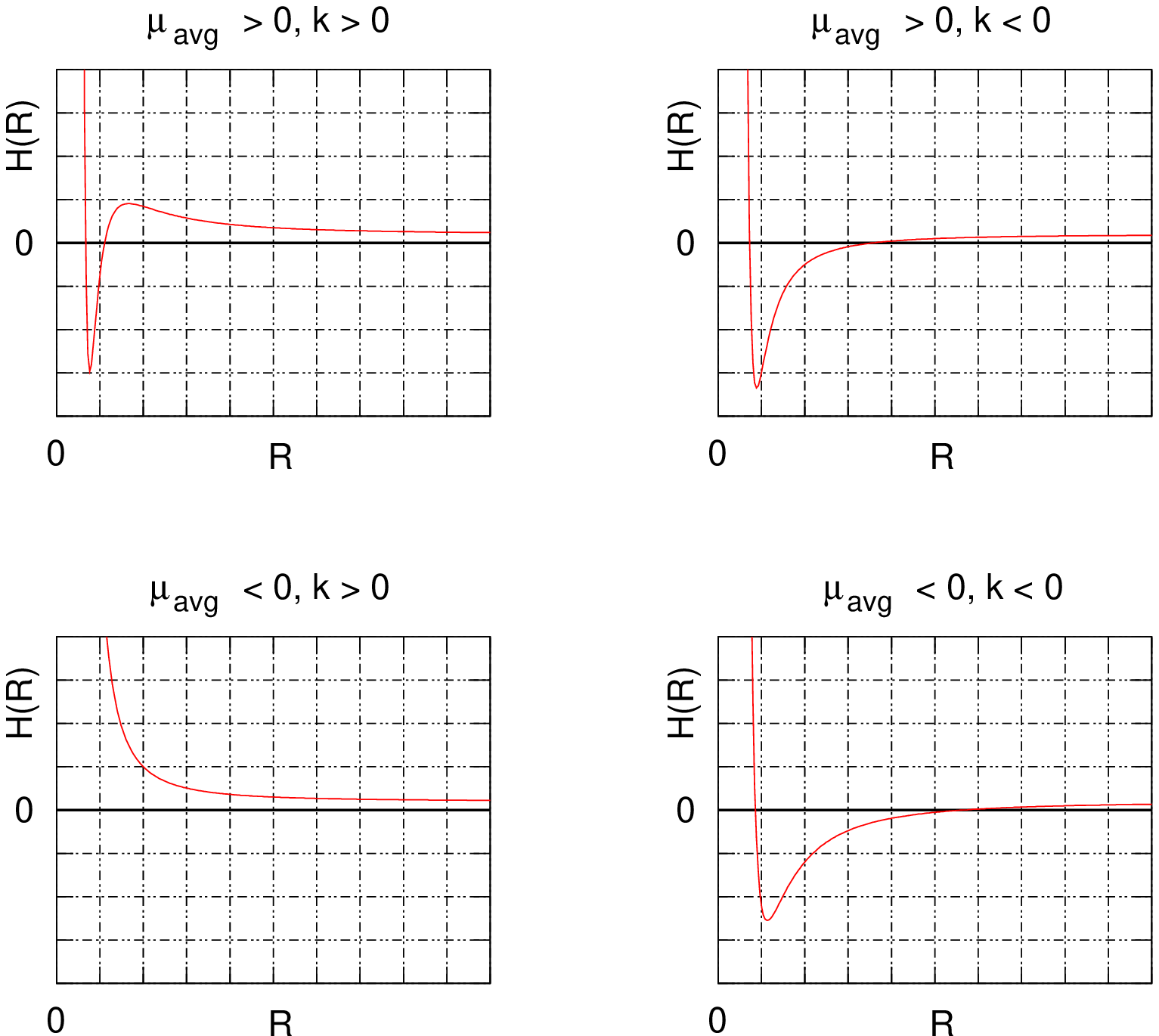}
  \parbox{12cm}{\vspace{.4cm}
    FIG.6 \hspace{2pt} The dependence of the Hubble constant on the scale factor (redshift) for different 
signs of the curvature and black hole mass\vspace{12pt}}
\end{center}

\section{Non-returning geodesics}
\label{divergences}

We saw that geodesics for which $F\left(r_{min}\right)=0$ and $\left.\frac{dF}{dr}\right|_{r=r_{min}}=0$ 
will not return to the brane. We show here that this condition is satisfied only for a particular geodesic:
the condition $\left.\frac{dF}{dr}\right|_{r=r_{min}}=0$ fixes the value of $r_{min}$ as a function of 
the curvature and bulk black hole mass:
\be
r_{min}^2=\frac{2\mu}{k}
\ee
Substituting this result in the equation $F\left(r_{min}\right)=0$, we obtain:
\be
F\left(r_{min}\right)=1-\frac{P^2}{E^2}\left(\frac1{l^2}-\frac{k^2}{4\mu}\right)=0
\ee
which fixes the value of the parameter $P^2/E^2$ characterizing the geodesic. We see that this condition 
selects a single geodesic, and in general this condition is not satisfied. Consequently, most of the 
geodesics will return to the brane.


\begin{references}
\bibitem{RS1}L. Randall and R. Sundrum, {\sl A Large Mass Hierarchy from a Small Extra Dimension},
Phys. Rev. Lett. {\bf 83} (1999) 4690, hep-ph/9905221.

\bibitem{RS2}L. Randall and R. Sundrum, {\sl An Alternative to
Compactification}, Phys. Rev. Lett. {\bf 83}, 3370 (1999), hep-th/9906064. 

\bibitem{Csaki_0}C. Csaki, M. Graesser, C. Kolda and J. Terning,
{\sl Cosmology of One Extra Dimension with Localized Gravity},
Phys. Lett. {\bf B462} (1999) 34, hep-ph/9906513.

\bibitem{Binetruy}P. Binetruy, C. Deffayet and D. Langlois,
{\sl Non-conventional cosmology from a brane-universe},
Nucl. Phys. {\bf B565} (2000) 269, hep-th/9905012.

\bibitem{Eanna}E. E. Flanagan, S.-H. H. Tye and I. Wasserman,
{\sl Cosmological Expansion in the Randall-Sundrum Brane World Scenario},
Phys. Rev. {\bf D62} (2000) 044039, hep-ph/9910498.

\bibitem{Ellwanger}P. Binetruy, C. Deffayet, U. Ellwanger and D. Langlois,
{\sl Brane cosmological evolution in a bulk with cosmological constant},
Phys. Lett. {\bf B477} (2000) 285, hep-th/9910219.

\bibitem{io}H. Stoica, S.-H. H. Tye and I. Wasserman,
{ \sl Cosmology in the Randall-Sundrum Brane World Scenario},
Phys. Lett. {\bf B482}, (2000) 205-212, hep-th/0004126.

\bibitem{Gregory}P. Bowcock, C. Charmousis and R. Gregory,
{\sl General brane cosmologies and their global spacetime structure},
Class. Quant. Grav. {\bf 17}, (2000) 4745-4764,
hep-th/0007177.

\bibitem{Kraus}P. Kraus, { \sl Dynamics of Anti-de Sitter Domain Walls}
JHEP {\bf 9912}, (1999) 011, hep-th/9910149. 

\bibitem{Grojean}C. Grojean, F. Quevedo, G. Tasinato and I. Zavala,
{\sl Branes on Charged Dilatonic Backgrounds: Self-Tuning, Lorentz Violations and Cosmology},
JHEP {\bf 0108}, (2001) 005, hep-th/0106120. 

\bibitem{Csaki_1}C. Csaki, J. Erlich and C. Grojean,
{\sl The Cosmological Constant Problem in Brane--Worlds and Gravitational Lorentz Violations},
gr-qc/0105114. 

\bibitem{Csaki_2}C. Csaki, J. Erlich and C. Grojean,
{\sl Gravitational Lorentz Violations and Adjustment of the Cosmological Constant in 
Asymmetrically Warped Spacetimes},
Nucl. Phys. {\bf B604}, (2001) 312-342, hep-th/0012143. 

\bibitem{Caldwell}R. R. Caldwell and D. Langlois,
{\sl Shortcuts in the fifth dimension}, Phys. Lett. {\bf B511}, (2001) 129-135,
gr-qc/0103070. 

\bibitem{Halevi} G. K\"albermann and H. Halevi,
{\sl Nearness through an extra dimension},
gr-qc/9810083. 

\bibitem{Freese}D. J. H. Chung and K. Freese, 
{\sl Can Geodesics in Extra Dimensions Solve the Cosmological Horizon Problem?},
Phys. Rev. {\bf D62}, (2000) 063513, hep-ph/9910235.

\bibitem{Kalbermann}G. K\"albermann,
{\sl Communication through an extra dimension},
Int. J. Mod. Phys. {\bf A15}, (2000) 3197, gr-qc/9910063. 

\bibitem{Ishihara}H. Ishihara,
{\sl Causality of the brane universe},
Phys. Rev. Lett. {\bf 86}, (2001) 381, gr-qc/0007070.

\bibitem{Creminelli}P. Creminelli,
{\sl Holography of asymmetrically warped space-times}, hep-th/0111107.

\bibitem{Birmingham}D. Birmingham, {\sl Topological Black Holes in Anti-de Sitter Space},
Class. Quant. Grav. {\bf 16}, (1999) 1197, hep-th/9808032.

\bibitem{Kolb}D. J. H. Chung, E. W. Kolb and A. Riotto,
{\sl Extra dimensions present a new flatness problem},
hep-ph/0008126. 

\bibitem{Dubovsky}S. L. Dubovsky,
{\sl Tunneling into Extra Dimension and High-Energy Violation of Lorentz Invariance},
hep-th/0103205. 

\bibitem{Csaki_3}C. Csaki,
{\sl Asymmetrically Warped Spacetimes}, hep-th/0110269.

\end{references}
\end{document}